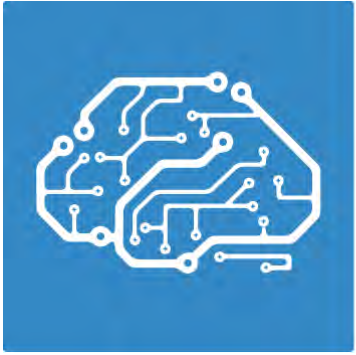

CBMM Memo No. 8     Apr-25-2014

# There's Waldo! A normalization model of visual search predicts single-trial human fixations in an object search task

by
Thomas Miconi, Laura Groomes, Gabriel Kreiman

Abstract:
When searching for an object in a scene, how does the brain decide where to look next? Theories of visual search suggest the existence of a global "attentional map", computed by integrating bottom-up visual information with top-down, target-specific signals. Where, when and how this integration is performed remains unclear. Here we describe a simple mechanistic model of visual search that is consistent with neurophysiological and neuroanatomical constraints, can localize target objects in complex scenes, and predicts single-trial human behavior in a search task among complex objects. This model posits that target-specific modulation is applied at every point of a retinotopic area selective for complex visual features and implements local normalization through divisive inhibition. The combination of multiplicative modulation and divisive normalization creates an attentional map in which aggregate activity at any location tracks the correlation between input and target features, with relative and controllable independence from bottom-up saliency. We first show that this model can localize objects in both composite images and natural scenes and demonstrate the importance of normalization for successful search. We next show that this model can predict human fixations on single trials, including error and target-absent trials. We argue that this simple model captures non-trivial properties of the attentional system that guides visual search in humans.

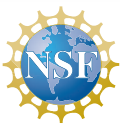

This work was supported by the Center for Brains, Minds and Machines (CBMM), funded by NSF STC award CCF-1231216.

**There's Waldo! A normalization model of visual search predicts single-trial human fixations in an object search task**

Abbreviated title:

There's Waldo! A normalization model of visual search

Authors: Thomas Miconi[1*], Laura Groomes[1], Gabriel Kreiman[1,2,3]

[1]Children's Hospital, Harvard Medical School

[2]Center for Brain Science, Harvard University

[3]Swartz Center for Theoretical Neuroscience, Harvard University

Number of pages: 32

Number of figures: 8

Number of words for:

Abstract: 210

Introduction: 652

Discussion: 634

The authors declare no competing financial interests.

Acknowledgements: This work was supported by NIH (GK) and NSF (GK).

Author contributions: TM designed the experiments, wrote the computational model, analyzed the data and wrote the manuscript. LG and TM collected the psychophysical data. GK discussed the model and experiment design and helped write the manuscript.

# Abstract

When searching for an object in a scene, how does the brain decide where to look next? Theories of visual search suggest the existence of a global "attentional map", computed by integrating bottom-up visual information with top-down, target-specific signals. Where, when and how this integration is performed remains unclear. Here we describe a simple mechanistic model of visual search that is consistent with neurophysiological and neuroanatomical constraints, can localize target objects in complex scenes, and predicts single-trial human behavior in a search task among complex objects. This model posits that target-specific modulation is applied at every point of a retinotopic area selective for complex visual features and implements local normalization through divisive inhibition. The combination of multiplicative modulation and divisive normalization creates an attentional map in which aggregate activity at any location tracks the correlation between input and target features, with relative and controllable independence from bottom-up saliency. We first show that this model can localize objects in both composite images and natural scenes and demonstrate the importance of normalization for successful search. We next show that this model can predict human fixations on single trials, including error and target-absent trials. We argue that this simple model captures non-trivial properties of the attentional system that guides visual search in humans.

# Introduction

Searching for an object in a crowded scene constitutes a challenging task. Yet, we can detect target objects significantly faster than would be expected by random search, even in a complex scene [1]. How does the brain identify the locations that might contain a target object? An influential concept suggests that the brain computes one or more ``priority maps'' which allocate a certain attentional value to every point in the visual space [2]. A large body of evidence shows that this attentional selection involves the Frontal Eye Field (FEF), the Lateral Intraparietal Area (LIP) and sub-cortical structures such as the Pulvinar and the Superior Colliculus (SC) [3-5]. How these areas interact with those involved in shape recognition is poorly understood.

In most models of visual search, the salience of an object is defined by the contrast between the object and its local surround along various features [2, 6-10]. Additionally, search behavior is influenced by the characteristics of the sought target [9, 11-18]. Top-down (task-dependent) guidance of visual search behavior occurs by increasing the weights of feature channels associated with the target (e.g. when looking for a red object, the red channel is given a higher weight in the computation of salience). Several models have combined bottom-up and top-down signals to deploy attentional modulation. Chikkerur and collages provided a derivation of attentional effects by applying Bayesian computations to image and target features [8] (see also [19, 20]). Navalpakkam and Itti extended the bottom-up saliency computation in [2] by introducing object-specific low-level feature biasing [7]. Small low-level local features (ON-OFF cells sensitive to intensity, orientation or color-opposition contrasts) are processed in independent maps with global competition, which are then combined in one global saliency map. Rao and colleagues used a multi-scale pyramid of simple filters to compute a signature of each point in the visual field and compared it with the signature of the searched object; visual search operates by repeatedly finding the point whose signature is most correlated with that of the searched object [9] (see also [14]). Hamker proposed a detailed, biologically motivated model of how attentional selection emerges from the reentrant interactions among various brain areas [10]. The model proposes that a semantic, target-specific top-down modulation, originating from the prefrontal cortex (PFC), falls upon object-selective cells in inferotemporal cortex (IT), and from there to visual area V4, raising neural responses to stimuli sharing features with the target. The ``oculomotor complex'' (FEF-LIP-SC) then aggregates activity from V4 into a global saliency map.

Motivated by recent advances in unraveling the mechanisms underlying visual search, here we introduce a model that describes how recognition computations interact with attentional selection. We set ourselves three specific requirements in developing this model. First, we seek a physiologically interpretable model based on plausible neural operations including multiplicative modulation and local normalization through divisive inhibition [5, 21]. Second, existing models are usually based on the assumption that target-specific modulation is applied to neurons in visual areas followed by integration into a spatial map within higher attentional areas [7, 8, 10, 22, 23]. However, recent observations suggest that target-specific modulation occurs in FEF before V4 or IT [24, 25] and appears to depend on the stimulus within the receptive field, rather than on neuron tuning [24, 26]. This suggests that target modulation is implemented via ``feature-guided'' spotlights, originating from attentional areas and affecting an entire cortical zone, rather than feature-based modulation of specific visual neurons. The model proposed here is constrained by these observations and seeks to describe how and where these feature-guided spotlights are computed. Third, we want to directly compare the predictions of the computational model against human behavior in a common visual search task involving complex objects in cluttered scenes.

We show that the proposed model can locate target objects in complex images. Furthermore, the model predicts human behavior during visual search, not only in overall performance, but also in single trials, including error and target-absent trials.

## Results

We consider the problem of localizing a target object in a cluttered scene (e.g. **Figure 2A-D**) and propose a computational model constrained by the architecture and neurophysiology of visual cortex (**Figure 1**). We start by describing the computational model, then discuss computational experiments demonstrating the model's ability to localize target objects and finally compare the model's performance with human psychophysics measurements.

**Model sketch and evaluation**

The model posits an "attentional map" modulated by features of the target object (**Figure 1; Methods**). This map is computed in a high-level retinotopic area, selective for relatively elaborate features, which we associate with macaque lateral intraparietal cortex (LIP, see Discussion). The key inputs to this area are (i) bottom-up signals from retinotopic shape-selective cells in earlier visual areas, and (ii) top-down signals that modulate the responses according to the identity of the target object. This top-down modulation involves target-specific multiplicative feedback on each cell in the area. This feedback input $F(\mathbf{O},P)$ (where $\mathbf{O}$ is the target object and P is the cell's preferred feature) is proportional to how much feature $P$ is present in the target object, and is learnt by exposure to images of the object (**Eq. 4;** left side of **Fig. 1**). In addition, the attentional map undergoes local normalization through divisive feedforward inhibition at every point (**Eq. 4**), effectively normalizing the bottom-up inputs. The attentional focus at any time is defined as the point of maximum local activity in the attentional map (**Eq. 6**).

To evaluate the performance of the model in localizing target objects in cluttered scenes we considered a set of 40 objects embedded in multi-object arrays (e.g. **Figure 2A,C**) or in natural images (e.g. **Figure 2B,D**). Examples of the attentional maps produced by the model for different input images and target objects are shown in **Figure 2E-H**. As expected, the attentional map depended on the target object and the background image. The attentional map showed above background activation in multiple locations that contain objects, with enhanced activation in the particular location where the target object was located both in composite images (**Figure 2E,G**) and natural background images (**Figure 2F, H**).

The model was deemed to have successfully localized the target if the attentional focus fell within the bounding box of the target object. If this was not the case, we implemented inhibition of return by subtracting a Gaussian-shaped field centered on the current maximum from the attentional map (**Eq. 7**). We then selected the new maximum of the resulting map as the second attentional target. To assess the performance of the model, we computed the proportion of successful localizations after 1 to 5 successive attentional fixations over all test images.

**Model performance**

We first applied our model to composite images, in which 9 objects were displayed on a blank background (**Figure 2A,C**). For each of 40 objects, we generated 40 images (total of 1600 images) comprising the object as a target and 8 other randomly selected objects (as distractors). The model found the target object on the first fixation in 52% of the composite images (**Figure 2I**, solid line) whereas randomly selecting one out of 9 positions would yield a success rate of 11.1%. For 90% of the images, the model found the target within four attentional fixations. We compared these results with a version of the system in which the feedback weights $F(\mathbf{O},P)$ were taken from a randomly selected object for each image ("Random weights", **Figure 2I**). The performance of this null model (**Figure 2I**, dotted line) was well below performance of the full model, and was essentially similar to what would be expected from randomly fixating successive objects. As an alternative control for bottom-up cues, we considered the saliency model of Itti and Koch [2, 34]. This purely bottom-up, task-independent architecture selects parts of an image that attract attention due to local contrast in intensity, orientation and color. As expected, the performance of this bottom-up model was comparable to that of the random weights model (**Figure 2I**, dashed line).

We investigated whether erroneous model fixations were driven by similarity between the fixated object and the target. We plotted the average similarity between successively fixated objects and target, along various measures of similarity, excluding fixations to the actual target (**Figure 3A-F**). Non-target objects attended to during the first fixations were more similar to the target under many similarity measures, including Euclidean distance (**3A**), pixel wise correlation (**3B**), correlation between luminance histograms (**3C**), absolute difference between mean luminance (**3D**) or size (**3E**) values, and Euclidean distance between C2b vectors produced by the bottom-up architecture (**3F**). We also evaluated whether certain object features correlated with ease of detection independently of the target. Object size (defined as the number of non-background pixels within the object bounding box) significantly correlated with probability of first fixation success for both composite images (r=0.34, p=0.03) and natural-background images (r=0.54, p < 0.0003). Object contrast (measured as the variance of non-background pixels) did not correlate with first fixation success in either type of image (composite images: r=-0.16, p>0.3; natural backgrounds: r=0.07, p>0.65).

Next, we applied our model to the more challenging task of detecting small target objects embedded within natural scenes (**Figure 2B,D**). We generated 1600 images (40 target objects and 40 natural scenes per object) where the area of the target object was 1/16 of the area of the whole image. The model localized the target object on the first fixation in 40% of the images (**Figure 2J** solid line, 15% for the randomized model and 13 % for the saliency model). Performance reached 77% after 5 fixations. As expected, the model's performance in natural scenes was significantly below the performance in composite images (with 9 objects).

The normalization operation (divisive feedforward inhibition by local inputs, **Eq. 5**) played an important role in the model's performance. In the absence of normalization, the system's performance was strongly degraded (**Figure 2**, "No normalization"). In the absence of normalization, locations with higher local feature activity tended to dominate the attentional map over the task-dependent feedback modulation (**Figure 3G**). Normalization by local inputs made it easier for object-selective feedback to drive lower-activity locations to prominence, by prioritizing the match between the input signals and top-down modulation over absolute magnitude of local inputs – effectively turning multiplicative modulation into a correlation operation (**Figure 3H**; see **Discussion**).

**Comparison between human and model behavior in a common task**

We sought to compare the performance of the model against human subjects in the same task. We recruited 16 subjects to perform a visual search task where they had to localize a target as fast as possible among 5 distractors in composite images by making a saccade to the target (**Figure 4A**, **Methods**). In ~30% of the trials, the target was absent. We tracked eye movements and compared the subjects' fixations against the model's predictions.

Subjects were able to perform the task well above chance levels (individual subject data in **Figure 4B,** average results in **Figure 4C**, "+" symbols). The task was not trivial as evidenced by the fact that the subjects' first saccade was only 65% correct in target-present trials (randomly selecting an object and other degenerate strategies would yield a performance of 16.6%; perfect detection in the first fixation would yield a performance of 100%). Subjects were essentially at ceiling by the $3^{rd}$ fixation. As expected from **Figure 2I**, the model was also able to successfully localize the target objects (**Figures 4B-C**, circles). The close match between model and subject performance evident in **Figures 4B-C** is probably dependent on experimental parameters including object eccentricity, similarity and other variables. Nevertheless, it is clear that the task is not trivial, yet both humans and the model can successfully localize the target within a small (and similar) number of fixations.

The overall similarity between psychophysical and model performance (**Figures 4B-C**) does not imply direct processing homologies between subjects and model, since both could perform target identification through entirely different mechanisms. To further validate the model, we sought to determine whether the model can also predict fine-grained aspect of subject behavior, namely single-trial fixations, beyond what could be expected purely from performance in target localization.

To compute an upper bound for how well the model could predict subject behavior, we first evaluated the reproducibility of subject responses, both within and across subjects. For this purpose, the same set of stimuli was shown to two different subjects, using the same target and the same array of objects for each trial, but randomizing object positions within the arrays and temporal order of trials. In addition, most subjects (13 out of 16) also participated in a second session in which, unbeknownst to them, they saw the same set of stimuli as in their first session, again randomizing object position and temporal order of trials. These "repeated trials" allowed us to estimate the consistency of responses both for individual subjects (within-subject agreement) as well as between two subjects (between-subject agreement). For these analyses, we chose to concentrate on the first fixation because it is the only one for which symmetry is ensured (e.g. after the first saccade, array objects are at different distances from the gaze location).

Subjects showed a high degree of self-consistency, defined as the proportion of repeated trials where the subjects first fixated on the same object, both individually (**Figure 5A**) and on average (**Figure 6**, black bars). In target-present trials, being able to locate the target with probability $p$ above chance suffices to lead to above-chance self-consistency. We evaluated the degree of self-consistency expected purely from the overall performance as $p^2 + (1-p)^2/5$. Subjects showed a stronger degree of self-consistency than predicted from performance in individual target-present trials (**Figure 5A1**). Furthermore, subject responses showed significant self-consistency in target-absent trials (**Figsure 5A2, Figure 6B**), and in trials for which the first fixated object was not the target in both presentations ("error trials", **Figs. 5A3, Figure 6C**), conclusively showing that consistency was not due to ability to locate the target.

As expected, consistency between subjects was slightly below self-consistency (compare black versus dark gray bars in **Figure 6**; see also **Figure 5A** versus **Figure 5B**). Still, between-subject consistency was also well above chance. In other words, different subjects showed consistent first fixation behavior when searching for the same target among the same set of choice objects.

Target absent trials were randomly intermixed with target present trials and therefore subjects could not tell on their first fixation whether the target was present or not. On target-absent trials, the degree of consistency of first saccades in trials with identical choice objects was significantly higher when the target was the same in both trials compared to when the target was different, both within and between subjects (**Figure 8**). This confirmed that the subjects' fixations were predominantly guided by target identity even when the target was not present in the image array (as opposed to being driven purely by bottom-up, or other target-independent features of the objects). Consistency in target-absent trials with identical choice objects but different targets at each presentation was slightly, but significantly above the chance level of 1/6 (**Figure 8**), both within-subject ($p<0.001$) and between-subjects ($p=0.03$; signed rank test of median=1/6 across 13 and 8 values, respectively). This indicates a weak, but significant effect of bottom-up, target-independent features in guiding saccades during this task.

Next, we investigated whether the model could predict the subjects' first fixations, including those in target-absent and error trials. We found that all the observations about self-consistency and between-subject consistency were also reflected in the agreement between model and subject responses (**Figure 5C** and light gray bars in **Figure 6**). The model's fixations agreed with the subjects' fixations on a trial-by-trial basis when considering all trials (**Figure 6A**), target-absent trials (**Figure 6B**) and error trials (**Figure 6C**). When comparing the responses of each individual subject with the model, above chance model-subject consistency was observed in all 16 subjects when considering all trials (**Figure 5C1**), in 15/16 subjects for target-absent trials (**Figure 5C2**) and 7/16 subjects in error trials (**Figure 5C3**). Overall, model-subject agreement was weaker than, but qualitatively similar to between-subject agreement.

To further investigate the agreement between model and subject responses, we plotted confusion matrices for first fixation position, across all images and all target objects (**Figure 6D-F**; see **Methods)**. These 6x6 matrices indicate how often the model's first fixation fell on position *i* when the subject's first fixation was on position *j*. The diagonal values in these matrices were significantly higher than the non-diagonal values ($p < .001$ for all three matrices, Wilcoxon rank sum test between the diagonal and non-diagonal values), illustrating the single-trial agreement between model and subjects in all trials (**Figure 6D**), target-absent trials (**Figure 6E**) and error trials (**Figure 6F**). Individual confusion matrices for each subject are shown in **Figure 7**.

## Discussion

We have introduced a simple model to explain how visual features guide the deployment of attention during search (**Figure 1**). This model proposes that a retinotopic area computes an attentional map, through the interaction of bottom-up feature-selective cells with a top-down target-specific modulatory signal and local normalization. An implementation of this model can locate complex target objects embedded in natural images or multi-object arrays (**Figure 2**). The model's performance also shows a significant degree of concordance with human behavioral performance in a relatively difficult object search task (**Figure 4**). The single-trial agreement between model and subjects extends to trials where the target was absent or where both the model and the subjects made a mistake (**Figure 5-6**).

Model performance cannot be explained by an overly easy task, as shown by the fact that human performance was far from perfect (**Figure 4**). Also, subjects showed a highly significant degree of self-consistency and between-subject consistency (**Figure 5-6**) but this consistency was far from 100%, reflecting considerable trial-to-trial variability. The agreement between the model and behavioral performance is also significantly above chance but far from 100%. The degree of between-subject consistency bounds how well we can expect the model to perform: it seems unrealistic to expect a model to be a better predictor of human behavior than humans themselves. Thus model-subject agreement should be evaluated in comparison with between-subjects agreement, as shown in **Figures 5** and **6**.

This model is inspired and constrained by current physiological knowledge about the macaque visual system. To search for a given object, we use the pattern of activity at the highest stages of visual processing, represented by the activity of C2b cells (left panel in **Fig. 1**), which are meant to mimic the output of bottom-up visual information processing along the ventral visual stream [28, 37]. Target-specific information interacts with bottom-up responses in an attentional area that we associate with lateral intraparietal cortex (LIP) as described by **Eq. 5**. We associate this computation with LIP rather than frontal eye fields (FEF) because, FEF has low visual selectivity in the absence of attention [38] whereas some studies have suggested evidence for shape selectivity in LIP [39, 40]. There is significant evidence implicating LIP in the generation of the attentional map [3, 41]. The top-down modulatory signal is presumed to originate from object-selective cells in a higher area. Prefrontal cortex (especially ventral lateral PFC) is a strong candidate for the source of the target-specific modulatory signal, since it is known to encode target identity in visual memory experiments [42, 43]. PFC also receives connections from object-selective visual areas such as IT and is also connected to LIP [44].

Visual search requires computing the match between bottom-up input signals and target modulation. Our experiments suggest the importance of local normalization for successful search, by preventing objects or areas with high bottom-up activity across all features from controlling the attentional map (**Figure 3G-H**). **Eq. 5** resembles a normalized dot product, which measures the similarity between two vectors independently of their magnitude; this approximation would be exact (neglecting the constant top-down signal) if the denominator in **Eq. 5** was strictly equal to the Euclidean norm of local activity. The interaction between normalization and modulation thus solves the problem of "pay[ing] attention to stimuli because they are significant, not simply because they are intense" [45], adding a new function to the many known computational properties of normalization in cortex [46].

In conclusion, our results show that a physiologically motivated model can not only match human performance in visual search, but also predict the finer details of human behavior in single trials. In combination with other models purporting to explain the mechanisms of attentional effects in lower visual areas (e.g. [5, 47-50]), this model can provide a component of a global mechanistic understanding of attentional selection in the brain.

## Materials and Methods

**Ethics Statement**

All the psychophysics experiments (described below) were conducted under the subjects' consent according to the protocols approved by the Institutional Review Board.

**Bottom-Up Architecture**

The computational model builds upon the basic bottom-up architecture for visual recognition described in [27, 28], which is in turn an elaboration of previous feed-forward architectures (e.g. [29-31]). This model relies on an alternation between "simple" cells that compute the match of their inputs with a pre-defined pattern, and "complex" cells that return the maximum of their inputs selective for the same pattern but at slightly different positions and scales. Here we consider two layers of simple and complex cells (S1, C1, S2b, C2b, using the same nomenclature as in previous studies) as described below.

We succinctly describe the bottom-up architecture here (for further details see [27, 28, 32]). We consider 256x256 pixel grayscale images $I(x,y)$ ($1 \leq x \leq 256$, $1 \leq y \leq 256$ pixels, $0 \leq I(x,y) \leq 255$). The first set of units (S1) convolve the image with Gabor filters at 12 scales $S$ ($S=1, 2... 12$) and 4 orientations $\theta$ ($\theta = 45, 90, 135, 180$ degrees). Following [28], the activation function for S cells is an L2-normalized inner product between weights and inputs. One difference with previous implementations is that we only use 12 different scales (rather than 16), and do not merge scales at any point: the model is essentially replicated in parallel at all scales. There are also minor differences in the positioning and spacing of cells at successive layers; in particular, the S1 cells do not densely cover the input. These choices result from early experimentation in which these particular arrangements provided a good trade-off between performance and speed.

Filters at scale $S$ are square matrices of size $D*D$, with $D = 7 + 2*(S-1)$ pixels. S1 cells are evenly spaced every $D/4$ pixels both vertically and horizontally - thus they do not densely cover the image. We enforce complete receptive fields (RF), which means that a cell's RF cannot overlap the border of its input layer. Because we enforce complete RF, the RF of the top-left-most cell is centered above pixel position $x=D/2, y=D/2$. Note that because of difference in RF diameters (and thus in margin and spacing), S1 cell columns of different scales do not generally fall at the same positions. At any given position, there is either no cell at all or a full column of 4 S1 cells (one per orientation), all of the same scale. This also applies to C1 and S2b cells, replacing orientations with prototypes for S2b cells (see below).

The Gabor filter G'$_{S,\theta}$ of scale S and orientation θ is defined for every row $x$ and column $y$ as $G'_{S,\theta}(x,y) = \exp\left(-\frac{\hat{x}^2 + \gamma^2\hat{y}^2}{2\sigma^2}\right)\cos(2\pi\hat{x}/\lambda)$ [27] where $\hat{x} = x\cos\theta + y\sin\theta$, $\hat{y} = -x\sin\theta + y\cos\theta$, $\lambda = 0.8\ \sigma$, $\sigma = 0.0036D^2 + 0.35D + 0.18$, $\gamma = 0.3$. Note that $-D/2 \leq x \leq D/2$ and $-D/2 \leq y \leq D/2$. The filter weights are then set to 0 outside of a circle of diameter $D$:

$G'_{S,\theta}(x,y:\sqrt{x^2+y^2} > D/2) = 0$. Finally, the Gabor filters are normalized to unit norm:

$G_{S,\theta}(x,y) = G'_{S,\theta}/\sqrt{\sum_{x,y} G'_{S,\theta}(x,y)^2}$. For a given S1 cell of scale $S$, orientation $\theta$, centered at position $(x_c, y_c)$, the output is the absolute value of the normalized inner product between the (vectorized) corresponding Gabor filter and the portion of the input image falling within the cell's RF [28]:

$$S1_{S,\theta,x_c,y_c} = \frac{\left|\sum_{i,j} G_{S,\theta}(i,j)I(x_c+i, y_c+j)\right|}{\sqrt{\sum_{i,j} I(x_c+i, y_c+j)^2}} \quad \text{[Equation 1]}$$

C1 layers take S1 output as their inputs. The output of a C1 cell of scale $S$ and orientation $\theta$ is the maximum of S1 cells of identical orientation and scale, within the RF of this C1 cell. At any scale, C1 cells are positioned over *every other* S1 column of the same scale, both vertically and horizontally. Each C1 cell returns the maximum value of all S1 cells of similar scale and orientation within a square of 9*9 S1cells centered at the same position as this C1 cell:

$$C1_{S,\theta}(x_c, y_c) = MAX_{i,j}(S1_{S,\theta}(x_c+i, y_c+j)) \quad \text{[Equation 2]}$$

with $-4 \leq i \leq 4$, $-4 \leq j \leq 4$. In the previous equation, $x_c$ and $y_c$ refer to position within the S1 layer of scale $S$, not to image pixel positions.

S2b cells take C1 output as their inputs. The output of an S2b cell depends on the similarity of its inputs with its prototype $P_S(i,j,\theta)$. There are 600 different prototypes, each of which takes the form of a 9*9*4 matrix as described below (9*9 diameter and 4 orientations). The same 600 prototypes are used for all scales. The output of an S2b cell of scale *S*, prototype *P* and position *x,y* is calculated as follows:

$$S2b_{S,P}(x,y) = \frac{\sum_{i,j,\theta} P_S(i,j,\theta) C1_{S,\theta}(x+i,y+j)}{\sqrt{\sum_{i,j,\theta} P_S(i,j,\theta)^2} \sqrt{\sum_{i,j,\theta} C1_{S,\theta}(x+i,y+j)^2} + 0.5} \qquad \text{[Equation 3]}$$

with $-4 \leq i \leq 4$, $-4 \leq j \leq 4$, and $\theta$ ranges over all 4 orientations. Note that the numerator describes a convolution of the entire stack of 4 C1 maps (one per orientation) with the S2b prototype, while the denominator normalizes this output by the norms of the prototype weights and of the inputs. Coordinates *x* and *y* refer to positions within the C1 grid of scale S, rather than image pixel positions. Following [27], each prototype $P_S(i,j,\theta)$ was generated by running the model up to level C1 on a random image, and extracting a patch of size 9*9*4 (diameter 9, 4 orientations) from the 4 C1 maps (one per orientation) at a random scale and a random location. Then, 100 randomly selected values from this patch were then kept unmodified, while all other values in the patch were set to zero. The resulting patch constituted the actual prototype P. This process was iterated until 600 prototypes were generated. The random images used to set the prototypes were distinct from all the images used in the computational experiments below (i.e. none of the 40 target objects or 250 natural images used under "Computational experiments" were used to determine $P_S(i,j,\theta)$). Note that the same set of 600 prototypes was used at all scales. The C2b layer returns the global maximum of all S2b cells of any given prototype *P*, across all positions and scales. Thus, there are 600 C2b cells, one for each S2b prototype *P*. The max operation in C2b as well as that in Equation 2 provide tolerance to scale and position changes [32].

**Attentional Selection**

The attentional model considers a situation where we search for object **O** in a complex image *I* that may contain an array of objects or a natural background (e.g. **Figure 2**). To search for object **O**, the model considers the responses of the bottom-up model to that object when presented in isolation: *C2b(***O***,P)* (1≤P≤600) and uses those responses to modulate the bottom-up signals to image *I*. We refer to this modulation as a feedback signal $F(\mathbf{O},P)$, defined by the normalized C2b output for the target object presented in isolation on a blank background:

$$F(\mathbf{O},P) = C2b(\mathbf{O},P) / \overline{C2b(P)} \qquad [\textbf{Equation 4}]$$

where $\overline{C2b(P)}$ is the average value of the C2b output for prototype *P* over 250 unrelated natural images. The dimension of *F* is given by the number of prototypes (600 in our case). Thus, for each S2b prototype *P*, the value of the feedback modulation *F*(**O**, *P*) when searching for target object **O** is proportional to the maximum response of this prototype to object **O** in isolation, across all positions and scales. *F* is then scaled to the [1,2] range by subtracting the minimum, dividing by the maximum coefficient, and finally adding 1 to each coefficient. This ensures that the total range of weights is the same for all objects. We note that *F* is not hard-wired; it is task-dependent and varies according to the target object.

The value of the feedback signal may be interpreted in a Hebbian framework: F represents feedback from "object selective cells" in a higher area (possibly identified with PFC, see Discussion) that receive inputs from, and send feedback to, all S2B cells. Under Hebbian learning, the connection from any S2B cell to each object-selective cell will tend to be proportional to the activation of this S2B cell when the object is present, and therefore so will the strength of the feedback connection, which determines F when the object-specific cell is activated during search. The learning phase for the F(**O**,P) weights is described in the left side of **Figure 1**.

The attentional model combines these target-specific signals with the responses of the bottom-up architecture up to S2b into a so-called LIP map (see Discussion), $\text{LIP}_{S,P,\mathbf{O}}(x,y)$, defined by:

$$\text{LIP}_{S,P,\mathbf{O}}(x,y) = \frac{\text{S2b}_{S,P}(x,y) * F(\mathbf{O},P)}{\sum_{k=1}^{k=600} \text{S2b}_{S,k}(x,y) + 5} \qquad \text{[Equation 5]}$$

At every position (*x, y*) in the LIP map (which correspond to positions in the S2b map), each LIP cell (of scale S and preferred prototype P) multiplies its S2b input by the attentional coefficient *F* for prototype *P* given the target object **O**. The denominator indicates that LIP cells also receive divisive feed-foward inhibition, equal to the sum of all incoming S2b inputs at this position. An additive constant in the denominator (corresponding to the "sigma" constant in the canonical normalization equation [33]) defines the "strength"' of normalization: a large value means that the denominator is dominated by the fixed constant and thus less dependent on local activity, while a low value means that the denominator is dominated by the variable, activity-dependent term. We empirically set this parameter to 5 for all simulations. As described below, divisive normalization is crucial to the performance of the model (see **Figure 3G-H** and Discussion).

The final attentional map used to determine the location of attentional focus, $A_{\mathbf{O}}(x,y)$, is simply defined as the summed activity of all LIP cells at any given position:

$$A_{\mathbf{O}}(x,y) = \sum_{S,P} \text{LIP}_{S,P,\mathbf{O}}(x,y) \qquad \text{[Equation 6]}$$

At any time, the global maximum of this attentional map defines the current fixation/attentional focus. Notice that this map is only defined at discrete positions of the original image - those over which S2b cells are located.

Due to their discrete support and divisive normalization (which compresses responses at the higher end of the range), the attentional maps produced by the system are difficult to interpret visually. For visualization purposes *only*, these maps are contrast-enhanced by linearly scaling them within the [0.5,1.5] range, then exponentiating the value of each point three times (x = exp(exp(exp(x)))); they are then smoothed by filtering the entire map with a Gaussian filter of standard deviation 3 pixels. Importantly, this processing is only used to generate **Figures 1** and **2**, for the purpose of assisting visualization. All the computations and results in this paper are based on the original, unprocessed $A_\mathbf{O}(x,y)$ as defined in **Equation 6**.

**Object Search**

Searching for a target object **O** in a given image operates by iteratively finding the position of the maximum of the attentional map $A_\mathbf{O}(x,y)$ defined in **Equation 6** and checking whether this maximum falls within the bounding box of the target object. The bounding box $B_\mathbf{O}$ was defined as the smallest square encompassing all pixels of the object. If $\arg\max[A_\mathbf{O}(x,y)] \in B_\mathbf{O}$, then the target has been found. Otherwise, we apply an *inhibition-of-return* (IoR) field to the attentional map, decreasing its value around the location of this maximum as described below. We then pick the next maximum, corresponding to a new model fixation, and iterate the cycle until the target has been found or the maximum number of fixations (set to 5 unless otherwise noted) has been reached.

The IoR procedure multiplies the current attentional map at fixation *f* pointwise by an inverted 2D Gaussian $N(x_F, y_F, \sigma_{IoR})$ centered on the position of the current (unsuccessful) fixation $(x_F, y_F)$ and with standard deviation $\sigma_{IoR}$:

$$A_\mathbf{O}(x,y)[f+1] = A_\mathbf{O}(x,y)[f](1 - k * N(x_F, y_F, \sigma_{IoR})) \qquad \text{[Equation 7]}$$

In all simulations we empirically set $k=0.2$ and $\sigma_{IoR}=16.667$.

We report the proportion of images where the target is found as a function of the number of fixations *f* required in the computational experiments in **Figures 2-4**.

**Control experiments**

We performed several controls and comparisons with other models. It is conceivable that in some cases, attentional selection could be driven by bottom-up "saliency" effects rather than target-specific top-down attentional modulation implemented via **Equations 5-6**. To evaluate this possibility, we compared the performance of the model as described above with two control conditions. First, we used a modified version of the model, in which attentional modulation used the weights ($F(\mathbf{O}', P)$) of a random object **O'** for every input image instead of the specific weights associated with the actual target object **O**. We refer to this control as "Random weights" in **Figures 2** and **4**. Second, we generated attentional maps based on the bottom-up saliency model of [2]. We used the Saliency Toolbox implemented in [34] with default parameters, except for setting 'normtype' to 'none' (using the default value for normtype results in very sparse saliency maps in which only a few of the objects have a non-zero saliency, leading to an even worse performance). We refer to this condition as "Saliency model" in **Figures 2** and **4**. Both control models were applied to the exact same images as the model and following the same procedure outlined under "Object search" above.

**Image similarity**

In **Figure 3**, we compared fixated objects and target objects. For this figure, we considered several possible similarity metrics: Euclidean distance (**3A**), pixel wise correlation (**3B**), correlation between luminance histograms (**3C**), absolute difference between mean luminance values (**3D**), absolute size difference (**3E**) and Euclidean distance between C2b vectors produced by the bottom-up architecture (**3F**).

**Computational Experiments**

We used the same set of 40 different target objects taken from [35] for all experiments. We also used natural images from [36]. All input images were grayscale squares of size 256*256 pixels. We tested the model's performance in locating target objects in two types of cluttered images: (i) composite images and (ii) natural backgrounds. Composite images consisted of a fixed number of objects (n=6 or n=9) on a blank background and were generated as follows. Nine objects, comprising the target object plus eight randomly selected objects different from the target, were resized to fit within a bounding box of size 43*43 pixels. This size results from seeking to maintain a margin of 20 pixels on all sides around each object, within a 256*256 image. These objects were then regularly placed over a uniform gray background (e.g. **Figure 2A,C**). For each of the 40 objects, we generated 40 images containing this object as the target, plus eight other randomly selected objects.

Natural background images consisted of a single target object (one of the 40 objects above resized to 64*64 pixels) superimposed onto one of 250 images of natural scenes. The insertion position was random (except to ensure that the entire object falls within the natural image). We considered 40 different images for each of the 40 target objects. Examples of these images are provided in **Figure 2B,D**.

For the learning phase during which the model learnt the appropriate weights for each particular object (left side in **Figure 1**; see above, **Attentional Selection**), we used images of the isolated objects, resized to a 64*64 pixel bounding box, on a uniform gray background.

**Psychophysics Experiments**

The task used for comparison between human and model behavior is illustrated in **Figure 4A**. We used the same set of 40 objects described above, resized to 156x156 pixels. The same object arrays were used for the psychophysics and computational model in **Figures 4-8**. Each object subtended ~5 degrees of visual angle. Stimuli were presented on a CRT monitor (Sony Trinitron Multiscan G520). We recruited 16 subjects (10 female, between 18 and 35 years old) for this experiment.

We used the Eyelink D1000 system (SR Research, Ontario, Canada) to track eye positions with a temporal resolution of 2 ms and a spatial resolution of ~1 degree of visual angle. We calibrated the device at the onset of each session by requiring subjects to fixate on visual stimuli located in different parts of the screen. The equipment was re-calibrated as needed during the experiment. A trial did not start if subjects' eyes were not within 1 degree of the fixation spot for a duration of 500 ms. Failure to detect fixation prompted for eye-tracking recalibration.

After correct fixation; subjects were shown a certain target object in the center of fixation for 1500 ms. The screen was then blanked for a 750 ms delay period. Then six objects were shown, regularly arranged in a circle around the center of the screen. The distance from the objects' center to fixation was ~8 degrees. The subject's task was to direct gaze towards the target object as quickly as possible. The subject was said to fixate a given object whenever its gaze fell within the bounding box of this object. If the subject's gaze found the target object, the trial ended and the object was surrounded with a white frame for 1000 ms in order to indicate success. If the target was not found within 1800 ms, the trial was aborted and a new trial began. Importantly, the target was present in 70% of the trials. The 30% of target-absent trials provided us with an additional point of comparison between human and model behavior (see text and **Figures 5-7**). Because these trials were randomly interspersed with the target present triasl, subjects could not tell whether the target was present or not without visual search. The consistency within and between subjects for target-absent trials show that subject response in these trials was not random, but was actually influenced by both target and array objects. Presentation order was randomized.

In each session, the subject was presented with a block of 440 trials (300 target-present, 140 target-absent). The same set of target objects and images within a block was shown to two different subjects (but randomizing the temporal order of trials and the position of objects in each trial). This allowed us to evaluate the reproducibility of behavioral responses between two subjects on single trials – that is, the proportion of trials in which both subject's first fixations were located on the same object (**Figures 5-7**).

In addition, 10 out of 16 subjects participated in a second session in which, unbeknownst to them, they were presented with the same block of trials as in their first session (again, randomizing temporal order of trials and object position within each trials). This allowed us to compute subject self-consistency (within-subject agreement).

It is conceivable that subject fixations could be driven or modulated by bottom-up factors, independently of target features. In the computational experiments in **Figure 2** we specifically compare the performance of our model against a purely bottom-up model (discussed above under "Control experiments"). For target-present trials, high performance in locating the target at first fixation indicates that target features do guide visual search. However, we do not have any such guarantee for target-absent trials, in which subjects might use a different, purely saliency-based strategy. To evaluate this possibility, the 140 target-absent trials in each block were composed of 70 different arrays of objects, each 6-object array shown twice, with a different object being the target each time (and randomizing the position of objects in the image). **Figure 8** shows that in target-absent trials, first fixations in two presentations of the same array of objects (randomizing object position) fall on the same object much more often when the target is the same across both presentations that when it is not. This observation shows that target features influence visual search even in target-absent trials.

**Confusion matrices**

We constructed confusion matrices to visualize the agreement between subjects and model across all single trials (**Figure 6D-F**). Each position *i, j* in these confusion matrices contains the number of trials for which the subjects first made a saccade towards position *i* and the model's maximum activation was on position j, divided by the number of trials in which the subjects first made a saccade towards position *i*. This represents the conditional probability that the model selected position *j*, given that the subject selected position *i*. These matrices summarize data across different objects. Because of the limited number of trials, we could not build a confusion matrix for each object at each position. Even though the object positions changed from trial to trial, we could still compare responses by subjects and model on a given trial by referring to these positions. The confusion matrices were computed separately for all trials (**Figure 6D**), target-absent trials (**Figure 6E**) and error trials in which both the subject and model chose an object different from the target (**Figure 6F**). The higher values along the diagonals in these matrices reflect the single-trial agreement between model and subjects (p < .001 for all three matrices, Wilcoxon rank sum test). In addition to these aggregate matrices, we also computed individual matrices for each subject, shown in **Figure 7**.

**FIGURE LEGENDS**

**Figure 1. Sketch of the attentional model**

Left panel: learning the attentional modulation coefficients *F(O,P)* for a given target object **O** (top hat). An image containing a target object (bottom) is fed to the bottom-up architecture of the model [28, 51] (**Methods**) involving a cascade of linear filtering with Gabor functions (S1) or image prototypes (S2b) and max operations (C1, C2b). Right panel: searching for object **O** in a given input image. The bottom-up output of S2b cells in response to a given image (bottom) is fed to an "LIP map", where it is modulated multiplicatively by the attentional signal *F(O,P)*, and divided by the total incoming S2b activity at each point. The total resulting activation at every point, summed across all scales and prototypes, constitutes the final attentional map $A_O(x,y)$ (top), corresponding to the output of the attentional system for guidance of attentional selection.

**Figure 2. Model performance in composite images and natural background images**

Example composite images (**A**,**C**) and natural background images (**B**,**D**) used for testing the model. The target object is a top hat in **A, B** and an accordion in **C, D**. **E-H**. Output of the model for each of the images in **A-D**. We show the attentional map ($A_O(x,y)$, top layer in **Figure 1**), after smoothing (see color map on the right, arbitrary units, **Methods**).

**I-J**. Performance of the model (asterisks) in locating 40 target objects in 40 composite images containing 9 objects (**I**) and 40 natural background images (**J**). For each possible number *x* of fixations ($1 \leq x \leq 5$)), the y-axis indicates the proportion of all images in which the target was found within the first *x* fixations. Error bars indicate standard error of the mean across all 40 objects. Dashed line with circles: model performance when the normalization step is omitted (no denominator in **Eq. 5**). Dotted lines with hexagons: model performance when feedback weights are randomly shuffled among objects for each input image. Dashed line with squares: attentional maps generated by a purely bottom-up saliency model that has no information about the target object [34]. The gray horizontal lines in **I** indicate increasing multiples of 1/9.

**Figure 3. Properties of the model**

**A-F**. Average similarity between target and successively fixated objects, using various measures of similarity or difference between images, excluding fixations to the actual target. The first bar in all graphs indicates the average similarity (or difference) between the target and the first fixated objects across all trials in which the first fixation was erroneous. The similarity or difference measures are: (**A**) pixel-wise Euclidean distance between the object images, (**B**) pixel-wise Pearson correlation between the images, (**C**) correlation between image luminance histograms, (**D**) absolute difference between mean luminance values (excluding background pixels), (**E**) absolute difference in size (i.e. number of non-background pixels within the bounding box) and (**F**) Euclidean distance between C2b vectors in response to the images. A significant correlation between fixation number and similarity or difference exists for all measures, except for size difference (**E**). The barely visible error bars indicate s.e.m. over the number of trials for each particular fixation; because we exclude fixations to the actual target, this number ranges from n=1158 (first column) to n=2997 (last column). These computational data are derived from the same images used for the psychophysics experiment (Figure 4), using target-present trials only.

**G-H**. Effect of normalization on model output. For each one of 40 objects, the x-axis indicates the average activity of all C2b cells elicited by the object. The y-axis indicates the average number of model fixations necessary to find the object in 40 composite images (if the object is not found after 5 fixations, the number of fixations is set to 6 for this image). Error bars are standard error of the mean over 600 C2b cells (horizontal) or 40 images (vertical). Without normalization (**G**), objects eliciting stronger C2b activity are easier to find, indicating that they attract attention at the detriment of other objects, biasing search (dashed line: $r$=-0.82, $p<10^{-11}$). With normalization (**H**), this effect disappears (r=-0.04, p=0.81).

**Figure 4. Comparing computer model and humans on a common task**

**A.** Visual search task. After fixation (verified by eye tracking, **Methods**) a target object was presented for 1500 ms. The screen was then blanked (except for a central fixation cross) for 750 ms, then a search array consisting of 6 objects was shown. If the subject failed to find the target after 2000 ms, the trial ends and a new trial began.

**B.** Comparison between model performance and individual subjects. Model performance ("o") versus subject performance ("+") on the same stimulus sets (same targets and same array of choice objects, but randomizing object positions within the array) for successive fixations for each individual subject. There are small variations for the model from one plot to another because model performance for each graph is estimated on the same stimulus set shown to the subject, which differs across subjects.

**C**. Average performance for subjects ("+", average of 16 subjects), model ("o"), and control models (random weights model shown with hexagons and saliency model shown with squares) for the task described in **A**. Only target-present trials averaged across subjects are shown here (see **Figure 5** for target-absent and error trials). Error bars indicate SEM across all 40 target objects. The two dashed lines represent the model performance when attentional weights were randomly shuffled across objects for each input image and from a purely bottom-up saliency model that had no information about the target object [34]. The horizontal dashed lines represent increasing multiples of 1/6.

**Figure 5. Consistency metrics for individual subjects**

We evaluated consistency in "repeated trials" where the same set of stimuli (identical target and same objects in different positions within array) was presented to two different subjects or two different sessions for the same subject. Within-subject agreement (**A1-3**): proportion of trials in which the subject first fixated the same object in repeated trials (13 subjects). Between-subject agreement (**B1-3**): proportion of repeated trials in which both subjects first fixated the same object (8 subject pairs). Model-subject agreement (**C1-3**): proportion of trials in which both the subject and the model first fixated the same object (16 subjects). Column 1 (**A1, B1, C1**) includes all trials. Column 2 (**A2, B2, C2**) includes only target-absent trials. Column 3 (**A3, B3, C3**) includes only error trials. In all plots, the dotted line indicates the chance level under the assumption of purely random (but non-repeating) fixations (1/6 in columns 1 and 2 and 1/5 in column 3). In the "all trials" case (column 1), we further consider a null model that takes into account the fact that subjects and model were able to locate the target above chance, which affects their level of agreement. If the two series being compared (same subject on repeated trials, two different subjects, or subject and model) have probability of finding the target at first fixation $p_1$ and $p_2$ respectively, then the probability of agreement by chance is $p_1 p_2 + (1-p_1)(1-p_2)/5$ ("+").

Whenever the degree of consistency was significantly above chance the comparison was marked with * ($p<0.05$, binomial cumulative distribution test).

**Figure 6. Consistency within subjects, across subjects, and between subjects and model**

**A-C**. Following the format and nomenclature in Figure 5, here we show average consistency values across subjects. Black: within-subject agreement (13 subjects); dark gray: between-subject agreement (8 subject pairs); light gray: model-subject agreement (16 subjects). Results are shown for all trials (**A**), target-absent trials (**B**), and target-present trials in which both responses were erroneous (error trials, **C**). Error bars indicate SEM across all subjects or pairs. The dashed line indicates chance performance (1/6 in **A**, **B** and 1/5 in **C**).

**D-F**. Subject-model confusion matrix for all trials (**D**), target-absent trials (**E**) and error trials (**D**). The color at row *i* and column *j* shows the conditional probability of the model's response (first saccade) being position *j* when the subject's response (first saccade) was position *i*. These matrices represent the average across all the objects (Methods); individual-specific matrices are shown in **Figure 7**. The color scales are different in the different panels (there was more consistency and hence a more pronounced diagonal between model and subject in correct target-present trials, which are part of **D** but not **E** or **F**; using the same scale would make it difficult to see the diagonal in **E** and **F**). Diagonal values are significantly higher than non-diagonal values for all three matrices ($p<0.01$, Wilcoxon rank sum test), reflecting the significant agreement between model and subject first fixations across trials.

**Figure 7**. Subject-model comparison for individual subjects. Individual confusion matrices for all 16 subjects, using all trials (**A**), target-absent trials (**B**) or error trials (**C**). The format for each confusion matrix is the same as that in **Figures 6D-F**. Matrices with diagonal values significantly higher than non-diagonal values indicate above-chance agreement between model and subject across trials (*: $p<0.05$, **: $p<0.01$, ***: $p<0.001$, n.s.: not significant, Wilcoxon

rank sum test).

**Figure 8**. **Target identity influences responses in target-absent trials.** Average self-consistency (**A**) and between-subject consistency (**B**) in responses to two trials where the target was absent, where the same 6-object array was shown (randomized object positions) and the target was either the same (same target) or different (different target). If the first fixation were driven purely by bottom-up signals derived from each object, we would expect similar degrees of consistency in the "same target" versus "different target" conditions. Instead, we observed a significantly higher consistency (both for within-subject comparisons ($p<10^{-5}$) as well as between subject comparisons ($p<10^{-4}$) when the target was the same, suggesting that subjects were using aspects of the target object to dictate their first fixation (Wilcoxon rank-sum tests with 13 and 8 pairs of values, respectively). Error bars indicate s.e.m. across 13 subjects for self-consistency, and 8 pairs of subjects for between-subject consistency. The horizontal dashed line indicates chance levels (1/6). Note that consistency in target-absent trials with different targets at each presentation is slightly, but significantly, above the chance level of 1/6, both within-subject ($p<0.001$) and between-subjects ($p=0.03$; signed rank test of median=1/6 across 13 and 8 values, respectively). This indicates a weak, but significant effect of bottom-up, target-independent features in guiding saccades in these images.

# Figure 1

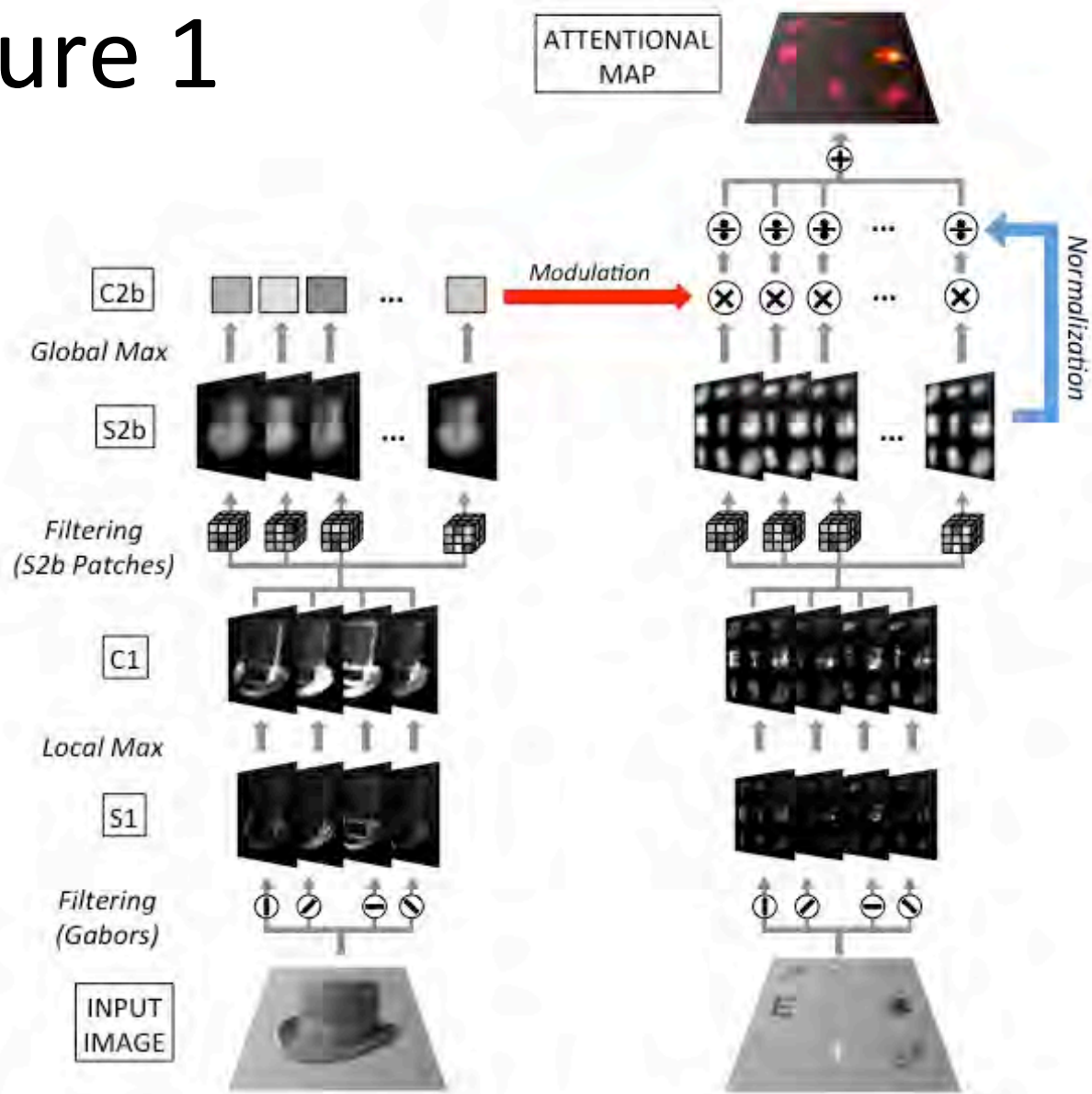

# Figure 2

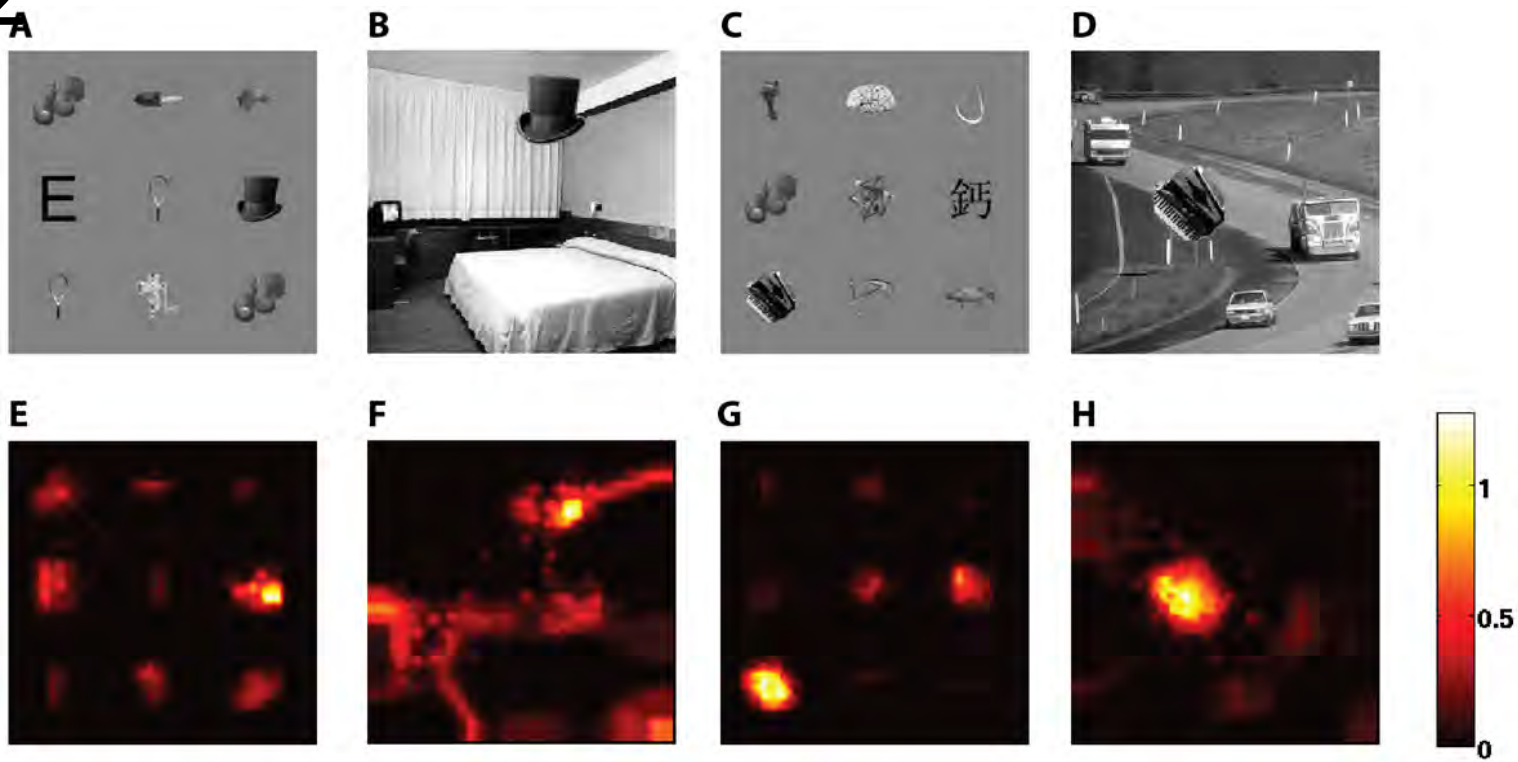

Figure 3

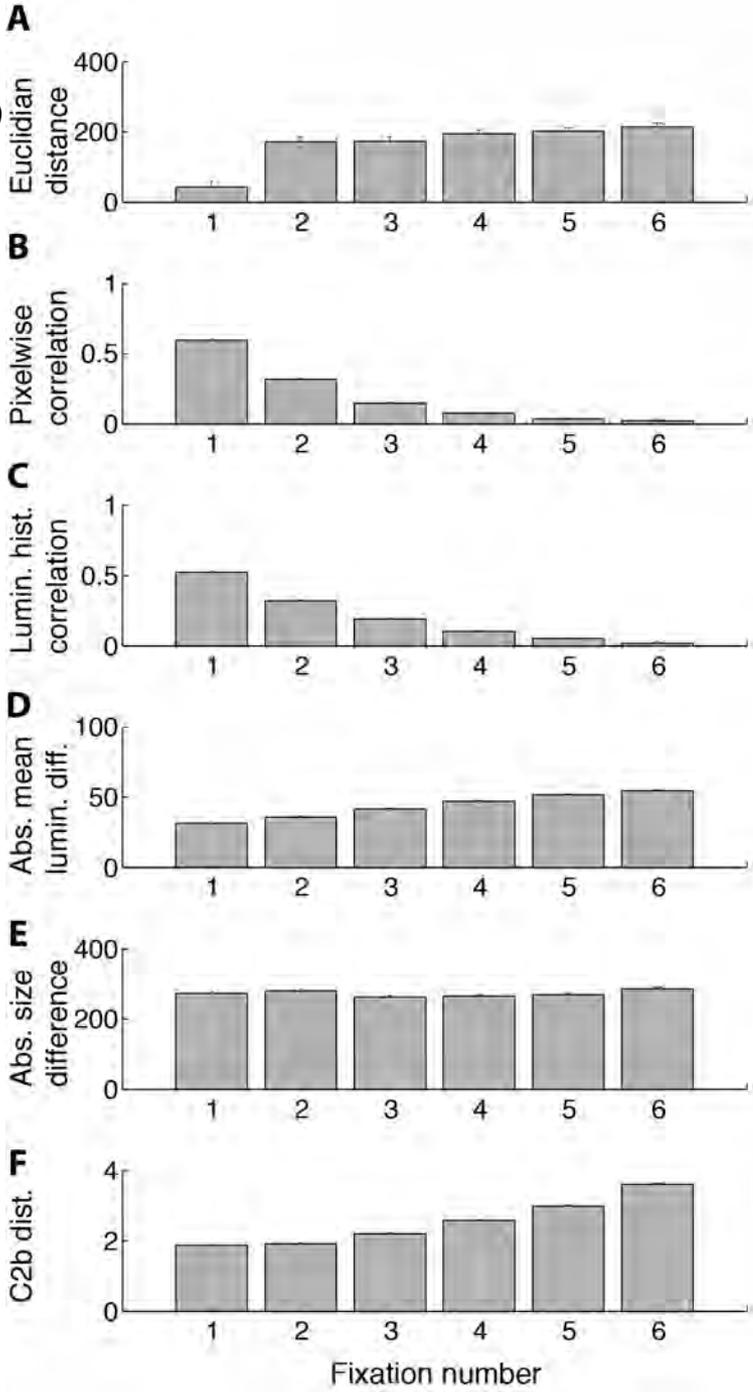
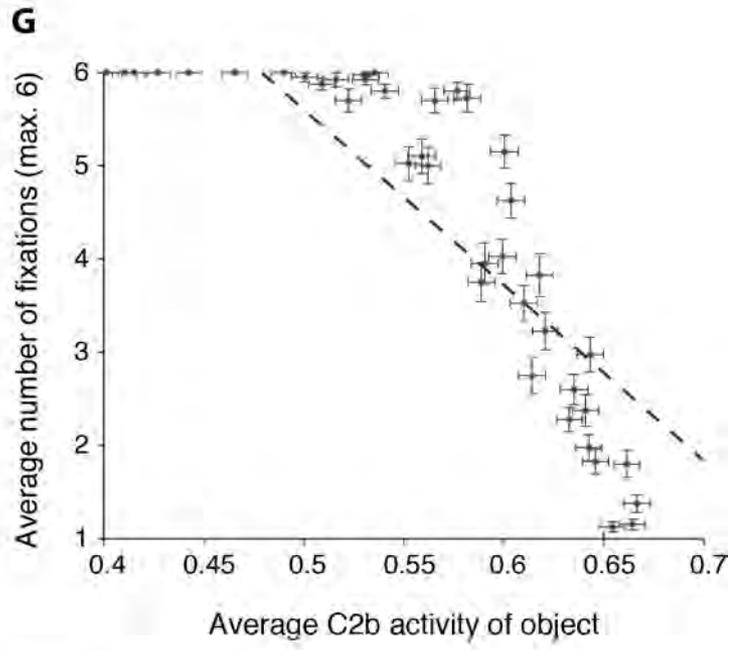
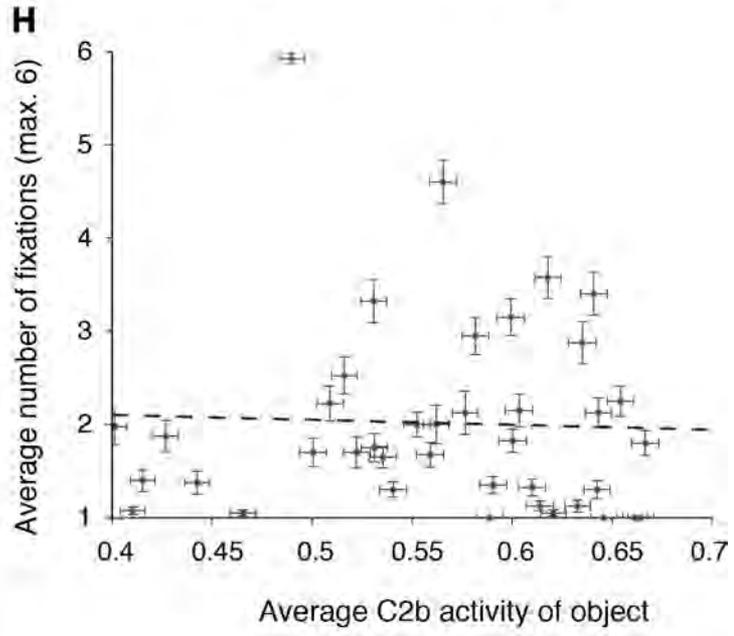

Figure 4

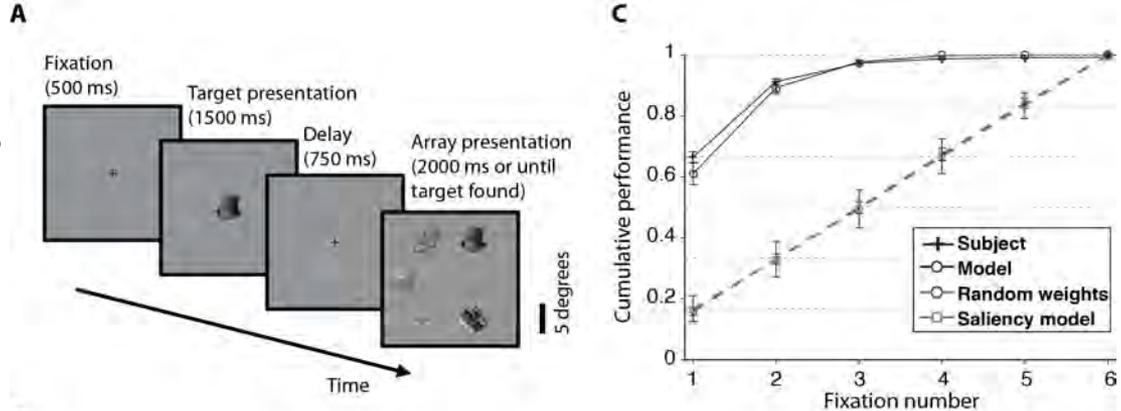

# Figure 5

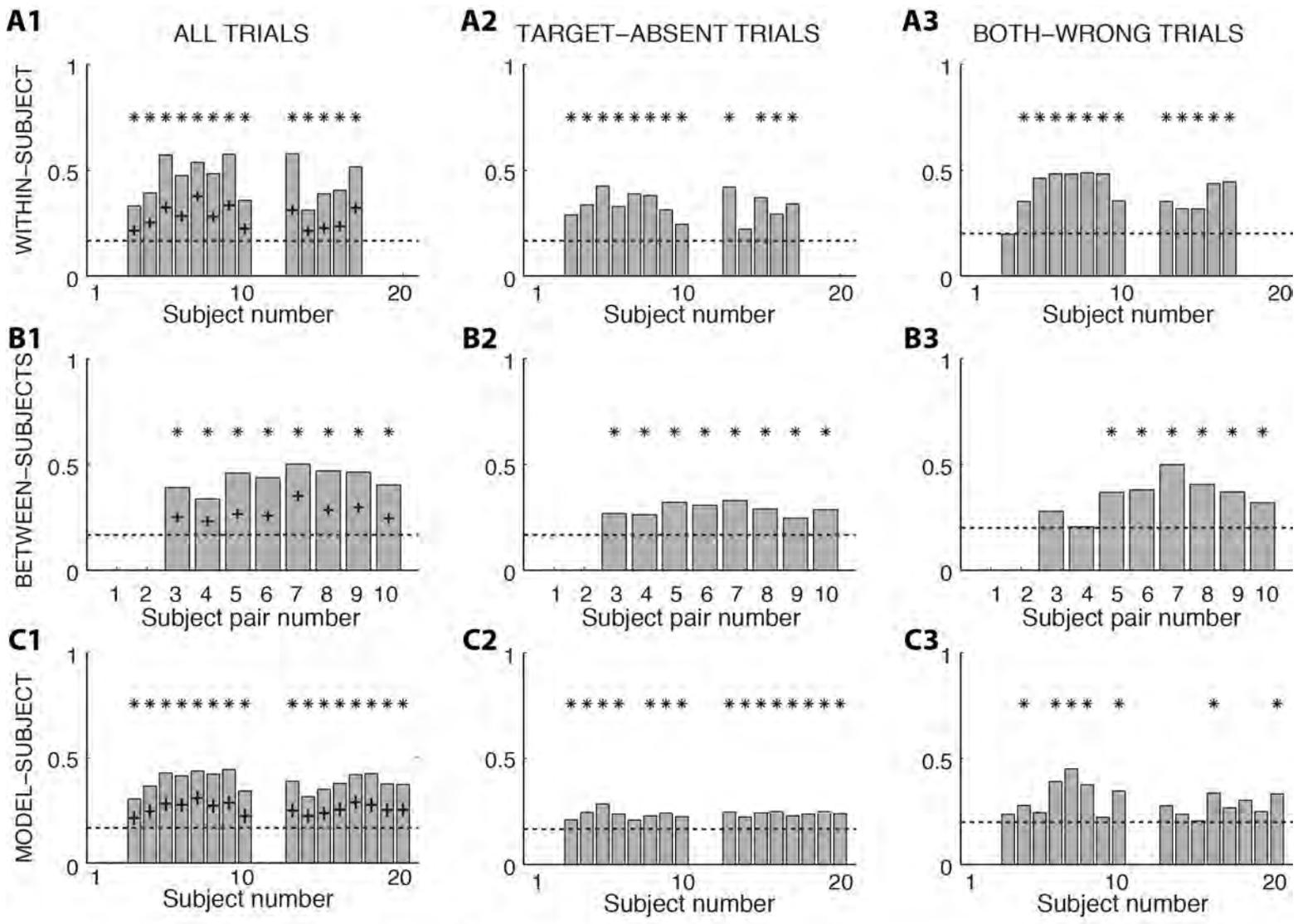

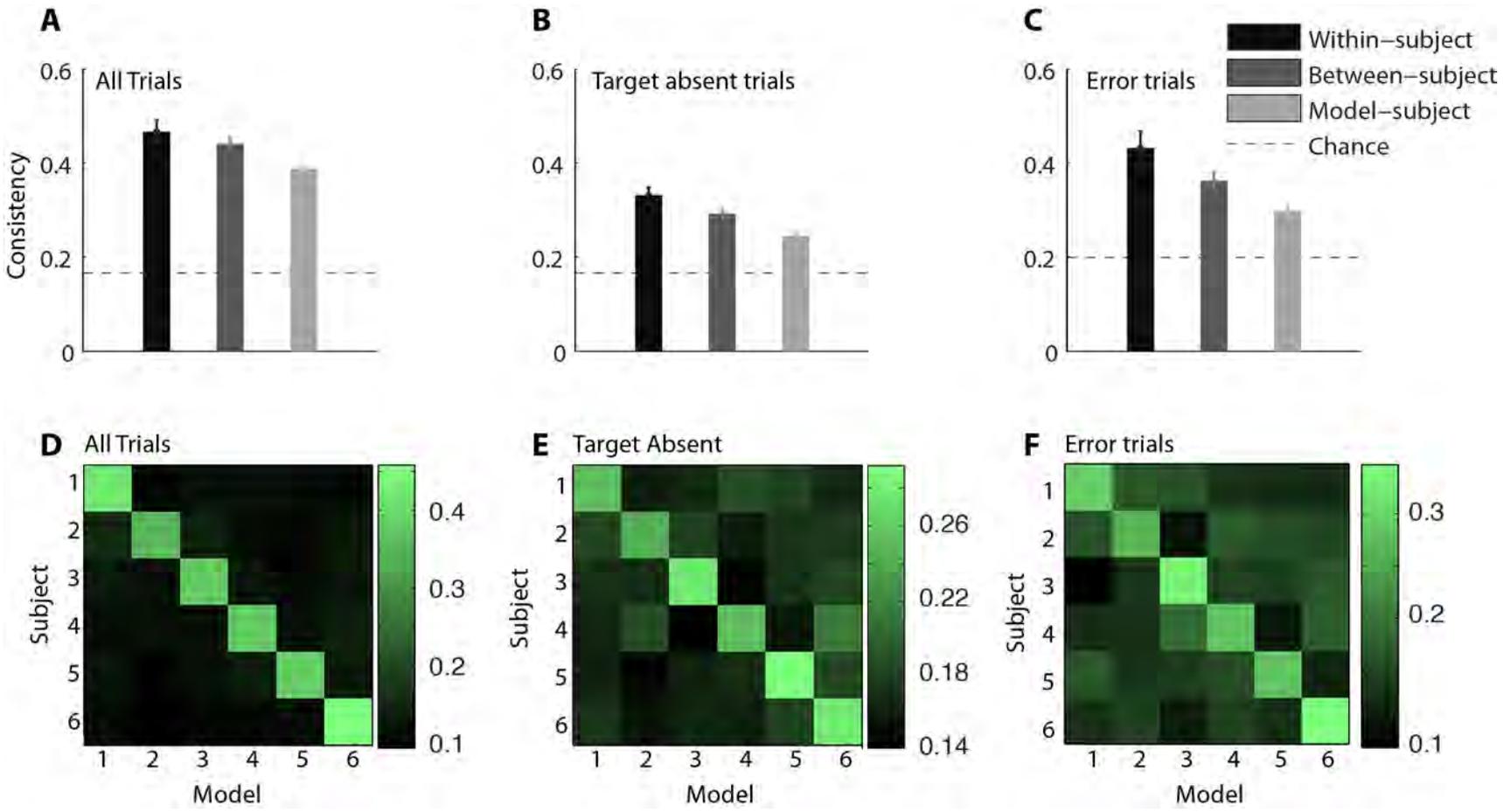

# Figure 7

**A** All Trials

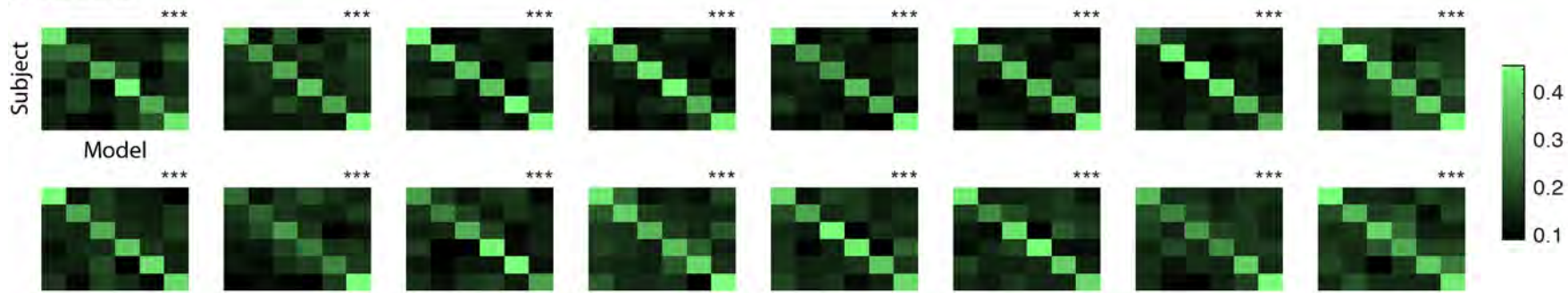

**B** Target Absent

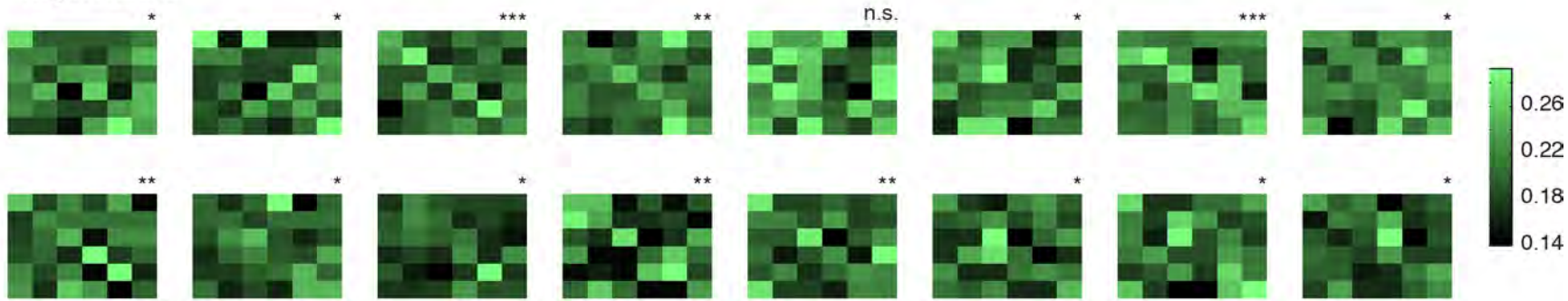

**C** Error trials

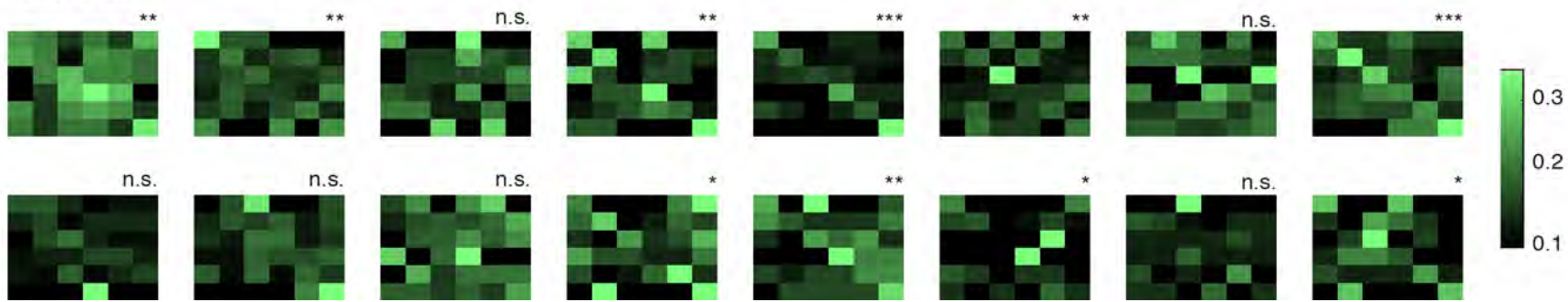

# Figure 8

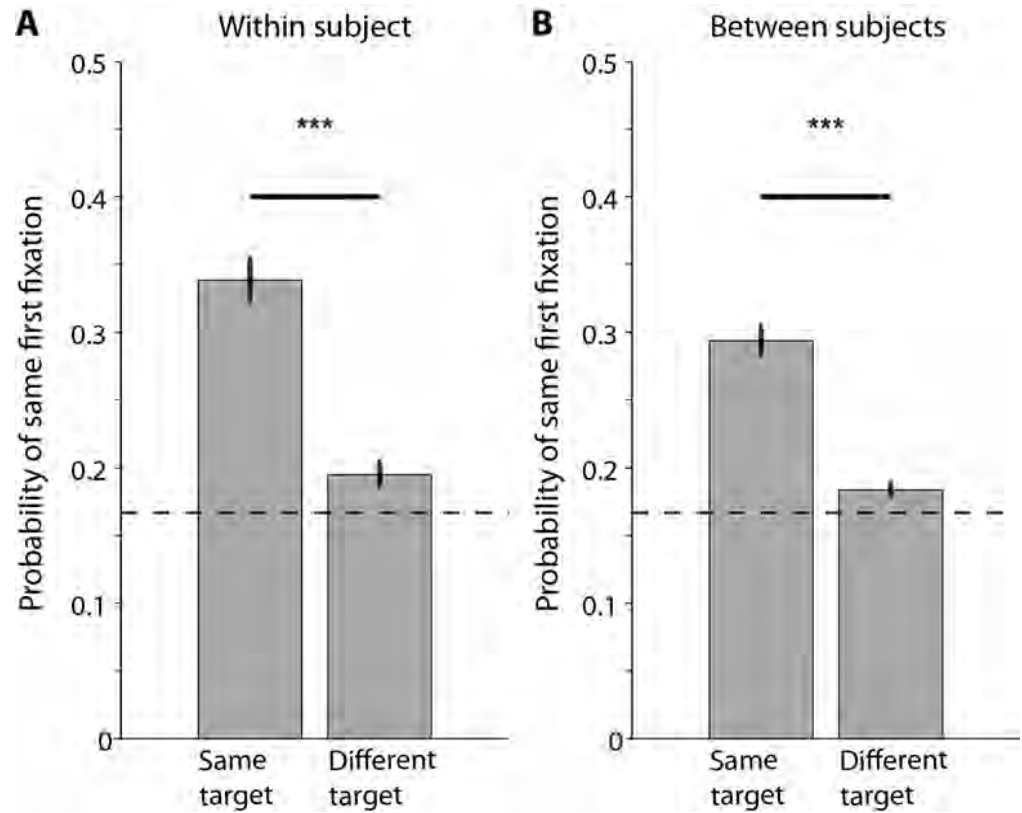